\documentclass[10pt, twocolumn]{article}

\usepackage{epsfig}
\usepackage{graphicx}
\usepackage{amsmath}
\usepackage{amssymb}
\usepackage{subfigure}
\usepackage{makecell}
\usepackage{array}
\usepackage{enumitem}
\usepackage[table,xcdraw]{xcolor}
\usepackage[normalem]{ulem}
\useunder{\uline}{\ul}{}
\usepackage[margin=1in]{geometry}

\usepackage{hyperref}
\hypersetup{
    colorlinks=true,
    linkcolor=blue,
    filecolor=magenta,      
    urlcolor=blue,
}
 
\title{\textbf{Studying the Impact of Mood on Identifying Smartphone Users}} 

\author{
  \textbf{K. Zanna\footnote{Khadija Zanna is a graduate student pursuing a Master's degree in Computer Engineering.}, S. King\footnote{Sayde King is a graduate student pursuing a Ph.D. in Computer Science and Engineering.}, T. Neal\footnote{Dr. Tempestt Neal is an Assistant Professor with research interests in behavioral biometrics and applied mobile sensing. She leads the Cyber Identity and Behavior Research Lab at USF. Homepage: \url{http://www.csee.usf.edu/\~tjneal/}}, and S. Canavan\footnote{Dr. Shaun Canavan is an Assistant Professor with research interests in affective computing and computer vision. He is a member of USF's Computer Vision and Pattern Recognition Group. Homepage: \url{http://www.csee.usf.edu/\~scanavan/}}} \\
  \small{Department of Computer Science and Engineering}  \\
  \small{University of South Florida, Tampa, FL USA 33620} \\
  \footnotesize{\texttt{\{khzanna,saydeking,tjneal,scanavan\}@mail.usf.edu}} \\ \\
  \footnotesize{\copyright 2019 The Authors}
}

\date{}

\begin{document}
\maketitle
\thispagestyle{empty}

\begin{abstract}
\small{\textit{This paper explores the identification of smartphone users when certain samples collected while the subject felt happy, upset or stressed were absent or present. We employ data from 19 subjects using the StudentLife dataset, a dataset collected by researchers at Dartmouth College that was originally collected to correlate behaviors characterized by smartphone usage patterns with changes in stress and academic performance. Although many previous works on behavioral biometrics have implied that mood is a source of intra-person variation which may impact biometric performance, our results contradict this assumption. Our findings show that performance worsens when removing samples that were generated when subjects may be happy, upset, or stressed. Thus, there is no indication that mood negatively impacts performance. However, we do find that changes existing in smartphone usage patterns may correlate with mood, including changes in locking, audio, location, calling, homescreen, and e-mail habits. Thus, we show that while mood is a source of intra-person variation, it may be an inaccurate assumption that biometric systems (particularly, mobile biometrics) are likely influenced by mood.}}
\end{abstract}

\section{Introduction}
Mobile biometrics is a growing research topic, with published literature ranging from continuous authentication using touch and keystroke data (e.g., \cite{185306, MENG20181}) to usability surveys on commercialized face and fingerprint systems (e.g. \cite{Bhagavatula15biometricauthentication, Javed:2017:IUC:3098954.3098974}). Behavioral mobile biometrics have gained attention considering the merits associated with transparent, continuous sensing compared to point-of-entry methods, particularly for user authentication \cite{7503170}. Specifically, point-of-entry methods may require nearly 50 unlock attempts per day \cite{185310}, cannot provide continuous authentication while using the device, and require users to remember complex combinations of characters or to present usable, physical biometric data. For some, point-of-entry methods have been regarded as awkward \cite{DeLuca:2015:IFL:2702123.2702141} or inconvenient \cite{185310}. 

Behavioral biometrics pose challenges as well, unfortunately. Behavioral signals, such as tone in voice \cite{Feng:2017:CAV:3117811.3117823},  touch location on the screen during swipes or flicks \cite{6331527}, and usage activity (e.g., app launches or calling patterns) may change over time due to various factors such as mood. Exactly how these changes affect mobile biometrics (particularly, usage activity) has yet to be clearly recognized. Meanwhile, the requirement that a biometric modality ``should be sufficiently invariant (with respect to the matching criterion) over a period of time'', or \textit{permanent}, is critical \cite{1262027}. Modalities that change often will have at least three main effects:
\begin{enumerate}[noitemsep, topsep=0pt]
    \item The template will need to be updated often to reduce false rejections \cite{Jain2011}.
    \item The hyperparameters of the matching algorithm may need to change over time to accommodate change in updated templates \cite{peel2015detecting}.
    \item Decision boundaries may move in feature space according to changes in which features are most important and/or as a consequence of changes to the matching algorithm.
\end{enumerate}

In this paper, we seek to understand the impact of mood on performance in a mobile biometric system. Summarizing our approach, we utilized data captured during the StudentLife study conducted by Wang et al. \cite{wang2014studentlife}. The StudentLife study employed a mobile app to continuously capture activity data such as GPS coordinates, app usage, and phone calls to later infer the impact of factors such as stress and activity on academic performance. Ecological Momentary Assessment (EMA) \cite{doi:10.1146/annurev.clinpsy.3.022806.091415} data was also captured via the StudentLife app, and used to label the activity data according to stress, number of hours slept, and mood, among others. EMA ``involves repeated sampling of subjects' current behaviors and experiences in real time, in subjects' natural environments'' \cite{doi:10.1146/annurev.clinpsy.3.022806.091415}, and can capture the dynamics of behavior in the real world \cite{asselbergs2016mobile}. In our study, we extracted features from the activity data and labeled each sample with student identifiers to re-evaluate the data from a biometric system perspective. To our knowledge, this is the first evaluation of the impact of mood on biometric performance using mobile activity data. We pose the following research questions: (1) \textit{How might feeling happy, upset, or stressed impact identification accuracy?}, and (2) \textit{How does feeling happy, upset, or stressed impact behavior which may consequently impact salient feature sets? }

The paper is outlined as follows. In Section \ref{background}, we discuss relevant background concerning mood, and the analysis of mood using mobile data. In Section \ref{data}, we describe the StudentLife dataset and extracted features. In Section \ref{method}, we detail our methodology; results are discussed in Section \ref{results}. Finally, we provide a brief discussion on limitations and conclude the paper in Section \ref{conclusion}. 

\section{Background} \label{background}
Mood is defined as ``affective states that are capable of influencing a broad array of potential responses, many of which seem quite unrelated to the mood-precipitating event. As compared with emotions, moods are typically less intense affective states and are thought to be involved in the instigation of self-regulatory processes'' \cite{morris}. Assessing mood is an important topic in psychology, and is largely done by analyzing physiological signals \cite{jaques2015predicting, sano2018identifying, sun2010activity}. With the widespread use of mobile devices, daily mood can now more easily be assessed through the use of these mobile devices \cite{ma2012daily}, suggesting a correlation between affect and mobile device usage. This correlation is further supported by years of research showing how mood affects decision making \cite{doi:10.1080/00049538908260083} and reasoning \cite{doi:10.2190/FYYA-GCRU-J124-Q3B2}, in conjunction with factors known to affect mood (e.g., weather \cite{denissen2008effects} and location \cite{doi:10.1111/j.1559-1816.1996.tb01781.x}), all of which play some role in the use of technology.

Few studies have evaluated the impact of mood on phone usage in the context of biometric recognition. One particular study evaluated the impact of stress on handwriting recognition \cite{BLANCOGONZALO20141173}, but most efforts have only \textit{assumed} that mood may affect usage which in turns affects recognition performance \cite{LiKamWa:2013:MBM:2462456.2464449}. However, several efforts have used phone data to recognize mood (e.g., \cite{Bardram:2012:MSS:2110363.2110370, Torous2015}), while others have shown that use of mobile devices themselves can induce certain moods \cite{doi:10.1556/2006.4.2015.010}.

LiKamWa et al. studied the mobile device activities of 32 subjects to infer mood \cite{LiKamWa:2013:MBM:2462456.2464449}. Their results demonstrated that the effect of mood on smartphone usage is person-specific using their \textit{MoodScope} sensing API. They trained least-squares multiple linear regression models per subject using app usage, phone calls, emails, SMS, web browsing, and location data. Achieving up to 93\% accuracy on classifying moods relevant to the Circumplex Mood Model (i.e., tense, stress, upset, bored, excited, happy, relaxed, and calm) \cite{russell1980circumplex}, this work provides valuable groundwork for understanding differences in device usage according to the user's affective state. In our work, we also base our analysis on the Circumplex model, examining happiness, stress, and upsetness. LiKamWa's work also suggests that certain behaviors (i.e., app usage and phone calls) are more indicative of mood as suggested by the results of feature selection using Sequential Forward Selection. 

In a similar study, Asselbergs et al. \cite{asselbergs2016mobile} conducted a pilot study using mobile phones for EMA. They collected text messages, screen time, app usage, accelerometer, and camera events from 27 subjects. Using this data, they created personalized models to predict daily variation in mood, accurately predicting up to 76\% of EMA mood scores. These results are encouraging, showing the feasibility of using mobile phones to predict \textit{person-specific} variations in mood, a key indicator that intra-person variation in mood may affect biometric performance.

More recently, Pratap et al. \cite{pratap2019accuracy} investigated the prediction of daily mood (e.g. depression) using phone data. Data was collected from 271 individuals over three months that included features such as call duration, SMS count, and SMS length. Using a random forest-based classifier, they too found that mood is more easily predicted based on individual subject models when using phone data. They showed that when predicting mood over all subjects, the prediction accuracy degraded. Similar to aforementioned studies, these results motivate our current research to better understand the impact of variations in mood on stability of phone usage when considering this data as a biometric modality. Specifically, we aim to further the analysis of mood by evaluating the impact of being happy, upset, and stressed on recognition performance. We view the possible impact of mood comparable to occlusions or pose in face recognition \cite{6468044}. By identifying \textit{mobile occlusions}, future efforts can work toward developing algorithms which can control or adapt to these factors.

\section{Data and Features} \label{data}
The StudentLife dataset was collected at Dartmouth College over a period of 10 weeks using an Android app running in the background \cite{wang2014studentlife}. It includes data from 48 students enrolled in a computer science programming class. Among the 48, 30 were undergraduates and 18 were graduate students. Thirty-eight subjects were male and 10 were female, consisting of 23 Asians, 23 Caucasians, and 2 African-Americans. 

The StudentLife data collection app required minimal interaction from participants, allowing continuous collection of mobile activities. A main requirement of the study was to carry the device throughout the day. The app collected data on activity, conversation, sleep, and location inferred from measurements from the accelerometer, microphone, light, GPS, and bluetooth sensors. The participants were also asked to respond to various EMA questions prompted at random to gather their self-reports of psychological health and academic performance. EMA prompts occurred eight times per day on average via the mobile app. A total of 35,295 EMA responses were collected, each requesting data on participant's current mood, stress, sleep duration, number of social interactions, exercise, and personality. EMA topics evaluated in this paper include stress, current mood, sleep quality, sadness, and happiness; their scales are provided in Table \ref{ema_scales}. Since each sample represented an hour of mobile activities (discussed below), we averaged the reported values for each EMA topic within the respective hour. However, we note that very few EMA responses were available and usually only a single EMA response was observed per hour.

\begin{table}[ht]
\centering
\caption{Ranges for EMA responses that are later used for classifying states of happiness, upsetness, and stress.}
\label{ema_scales}
\resizebox{\columnwidth}{!}{  
\renewcommand{\arraystretch}{1.3}
\begin{tabular}{l|l|l|l}
\hline
\textbf{Stress} & \textbf{Current Mood} & \textbf{Sleep Quality} & \textbf{Happy / Sad} \\ \hline
\begin{tabular}[c]{@{}l@{}}(1) Stressed out\\ (2) Definitely stressed\\ (3) A little stressed\\ (4) Feeling good\\ (5) Feeling great\end{tabular} & \begin{tabular}[c]{@{}l@{}}(1) Happy\\ (2) Stressed\\ (3) Tired\end{tabular} & \begin{tabular}[c]{@{}l@{}}(1) Very good\\ (2) Fairly good \\ (3) Fairly bad \\ (4) Very bad\end{tabular} & \begin{tabular}[c]{@{}l@{}}(1) A little bit \\ (2) Somewhat \\ (3) Very much \\ (4) Extremely\end{tabular} \\ \hline
\end{tabular}}
\end{table}

For this study, we utilized approximately 60 days of data from the StudentLife dataset collected from March 25, 2013 through May 25, 2013. Some subjects were more active in the use of the device and/or in providing EMA responses; 19 of 48 subjects were most active, and were therefore included in our analysis. Data from six mobile activities were used, including call, app, GPS, audio, activity, and locking events. We extracted 1,463 feature samples (one representative of each hour of mobile activity) of 7,834 dimensions per subject of the following features:
\begin{description}[noitemsep, topsep=0pt]
\item[Call] The sum of call durations that were started within the hour, resulting in a single call feature. Overall, 1,195 calls made across all subjects were less than 24 minutes, 34 were between 24 and 48 minutes, 8 were between 48 and 71 minutes, and 5 were between 71 and 119 minutes. One call lasted for 237 minutes, while another (which we considered an extreme anomality) was 2,386 minutes (nearly 40 hours). Many (26,553) samples did not have calling data.
\item[App] Each app is identified by a task ID. The task ID for every running app was extracted, and this list was reduced to a set (i.e., duplicates removed). Finally, app data across all users were transformed using one-hot encoding for consistency, resulting in 7,802 app features per sample.
\item[GPS] All latitude and longitude pairs were extracted, and then clustered using DBSCAN, a well-known density based clustering algorithm \cite{Ester:1996:DAD:3001460.3001507}, using the default parameters set by Scikit-Learn. DBSCAN clustering yielded 17 unique locations across all users; clustering labels were then used in lieu of latitude/longitude pairs. Finally, all labels across users were transformed using one-hot encoding for consistency, resulting in 28 GPS features per sample.
\item[Audio] Audio data were classified as (0) silence, (1) voice, (2) noise, or (3) unknown. We extracted the mode of the audio inferences; if multiple modes existed, we assigned a value 4. If no audio data existed, we assigned a value -1, overall resulting in a single audio feature per sample. 4,504 samples did not have audio data (-1), 17,061 samples were associated with silence, 3,549 samples were associated with the sound of a voice, 2,673 were associated with noise, and 10 had multiple audio inferences.
\item[Activity] Activity data were classified as (0) stationary, (1) walking, (2) running, and (3) unknown. We followed the same protocol as implemented for audio data, resulting in a single activity feature per sample. 4,506 samples did not have activity data (-1), 22,491 samples were associated with the subject being stationary, 365 samples were associated with walking, 110 were associated with running, and 322 samples had unknown activities. Only four samples had multiple activity modes.
\item[Lock] This feature represents the number of times the device was locked within the hour, resulting in a single lock feature. The number of locking events within an hour ranged from zero to four across the dataset. 21,425 samples had no locking events, 5,032 had one locking event, 1,332 had two locking events, 7 had three, and 1 had four locking events.
\end{description}

Of the 27,797 total samples, only 2,855 had corresponding EMA data for at least one of the five EMA topics (i.e., stress, current mood, sleep quality, happiness, and sadness), equating to only 10.3\% of the total dataset. To increase the number of samples with corresponding self-reports of mood, we inferred hourly and daily mood to create two additional datasets. To generate the second dataset, we assigned EMA values at hour $h$ to $h-1$ and $h+1$ for each user and each EMA topic if the samples at $h-1$ or $h+1$ had no EMA data. This increased the number of samples with corresponding EMA responses to 5,291 (19\%).  When referring to this dataset, we use $H$. Finally, we inferred the users' daily moods by averaging all EMA data per EMA topic for each day and assigning every sample within that day with this averaged response, increasing the number of samples with EMA data to 16,877 (61\%). When referring to this dataset, we use $D$. Our motivation for these approaches to consider the possibility of longer segments of mood stems from the fact that mood is known to persist longer than emotion \cite{ekkekakis2012affect, doi:10.1080/09515080600806567}. With datasets $H$ and $D$, we then assigned each sample a corresponding mood label of happy, upset, or stress according to the ranges of EMA values specified in Table \ref{ema_ranges}. 

\begin{table}[ht]
\centering
\caption{Ranges in EMA responses for establishing happy, upset, and stress.}
\label{ema_ranges}
\resizebox{\columnwidth}{!}{  
\renewcommand{\arraystretch}{1.3}
\begin{tabular}{c|c|c}
\hline
\textbf{Happiness} & \textbf{Upset} & \textbf{Stress} \\ \hline
\begin{tabular}[c]{@{}l@{}} Sleep quality = 1 \\ Stress level $\geq$ 4 \\ Happiness $\geq$ 2 \\ Current Mood = 1 \end{tabular} & 

\begin{tabular}[c]{@{}l@{}} Stress level = 3 \\ Current Mood = 3 \& Sleep quality $>=$ 3 \\ Sadness $\geq$ 2 \end{tabular} & 

\begin{tabular}[c]{@{}l@{}} Sleep quality $\geq$ 3 \\ 1 $\leq$ Stress level $\leq$ 3 \\ Current Mood = 2 \end{tabular} \\ \hline
\end{tabular}}
\end{table}

\section{Method} \label{method}
The goal of this work was to conduct identification experiments with an independent variable of mood; to accomplish this, we ran identification experiments without the mood of interest. In biometrics, identification seeks to establish an identity of an unknown individual by comparing that individual's physical or behavioral characteristics with those from several other candidates. Thus, identification requires a one-to-many match per subject, where the candidate with the most similar characteristics is returned as the identity of the unknown subject. We used a random forest classifier with 250 trees for our experiments, utilizing samples extracted from hour $h$ to $h+\delta$ for training from every subject, and samples extracted from hour $h+\delta+1$ to $h+2\delta+1$ for testing from every subject per hour; $\delta$ ranged from 4 to 24 (i.e., four hours to one day). For each train-test split, we performed feature selection using an Extra Trees classifier with 50 estimators. Thus, salient features may change over time, which we discuss in the following section.

\section{Results} \label{results}

To evaluate performance, we utilize the $F$-score, or the harmonic mean of precision and recall. If $TP$, $FP$, and $FN$ correspond with a true positive, false positive, and false negative, respectively, then the $F$-score, precision, and recall are defined as follows:
\[ F-score = 2 \times \frac{precision \times recall}{precision + recall} \]
\[ Precision = \frac{TP}{TP + FP} \]
\[ Recall = \frac{TP}{TP + FN} \]

The discussion in this section is based on the comparison of $F$-scores when excluding the samples corresponding with the mood of interest versus the use of all samples. Table \ref{tbl:samplespersubject} shows the number of mood-related samples per subject and the total number of samples removed when evaluating each mood. We do note that the number of mood-related samples per subject is imbalanced, especially for stress; this may bias the results, and we plan to consider alternative approaches to address this in future work.

\begin{table}[ht]
\centering
\label{tbl:samplespersubject}
\caption{Number of mood-related samples per subject.}
\resizebox{\columnwidth}{!}{  
\renewcommand{\arraystretch}{1.3}

\begin{tabular}{|l|l|l|l|l|l|l|}
\hline
 & \textbf{Happy ($H$)} & \textbf{Upset ($H$)} & \textbf{Stress ($H$)} & \textbf{Happy ($D$)} & \textbf{Upset ($D$)} & \textbf{Stress ($D$)} \\ \hline
1 & 346 & 389 & 452 & 934 & 574 & 1054 \\ \hline
2 & 216 & 159 & 296 & 838 & 504 & 1006 \\ \hline
3 & 190 & 145 & 265 & 696 & 480 & 960 \\ \hline
4 & 179 & 123 & 259 & 647 & 480 & 936 \\ \hline
5 & 137 & 120 & 234 & 598 & 384 & 912 \\ \hline
6 & 115 & 98 & 203 & 528 & 336 & 886 \\ \hline
7 & 77 & 96 & 202 & 336 & 334 & 789 \\ \hline
8 & 67 & 89 & 187 & 336 & 334 & 574 \\ \hline
9 & 48 & 68 & 143 & 288 & 312 & 552 \\ \hline
10 & 48 & 66 & 138 & 264 & 264 & 552 \\ \hline
11 & 48 & 60 & 128 & 216 & 192 & 552 \\ \hline
12 & 44 & 57 & 128 & 192 & 192 & 552 \\ \hline
13 & 44 & 44 & 111 & 168 & 192 & 504 \\ \hline
14 & 40 & 43 & 100 & 168 & 168 & 480 \\ \hline
15 & 34 & 32 & 84 & 168 & 168 & 360 \\ \hline
16 & 32 & 28 & 52 & 120 & 142 & 264 \\ \hline
17 & 28 & 24 & 48 & 120 & 120 & 238 \\ \hline
18 & 4 & 0 & 20 & 24 & 0 & 72 \\ \hline
19 & 4 & 0 & 0 & 24 & 0 & 0 \\ \hline
\textbf{Total} & \textbf{1701} & \textbf{1641} & \textbf{3050} & \textbf{6665} & \textbf{5176} & \textbf{11243} \\ \hline

\end{tabular}}
\end{table}

\subsection{Does mood impact identification performance?}

\begin{figure}[ht]
    \centering
     \includegraphics[width=\columnwidth]{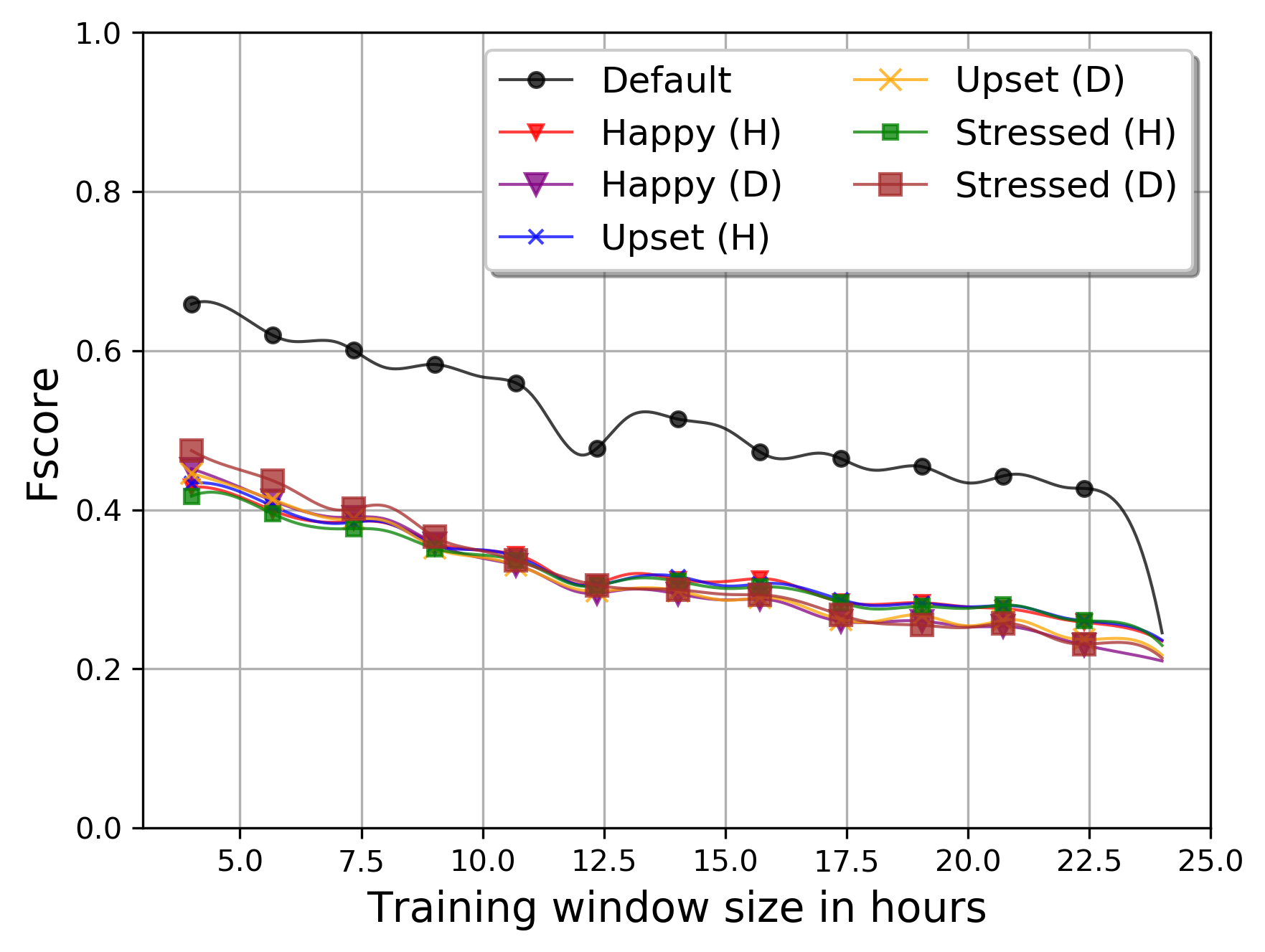}
    \caption{$F$-scores for each experiment that excludes mood-related samples across different training set sizes.}
    \label{fig:fscores}
\end{figure}

Figure \ref{fig:fscores} provides the ranges of $F$-scores as the training size increases from 4 to 24 hours. The default configuration, which is the use of all samples, outperforms all configurations that exclude those samples indicative of mood. Interestingly, our findings show that the three moods impact performance very similarly when those samples are removed, and there is little evidence supporting the need to exclude certain data based on mood. Thus, while many works have claimed that mood may have a negative impact on identification performance, our findings show the inverse. This assumption is typically made considering the variability in behavior induced by mood, which may correspond with decreased intra-person similarity. To evaluate this, we examine the features chosen during feature selection in the following section.

\begin{figure}[ht]
    \centering
     \includegraphics[width=\columnwidth]{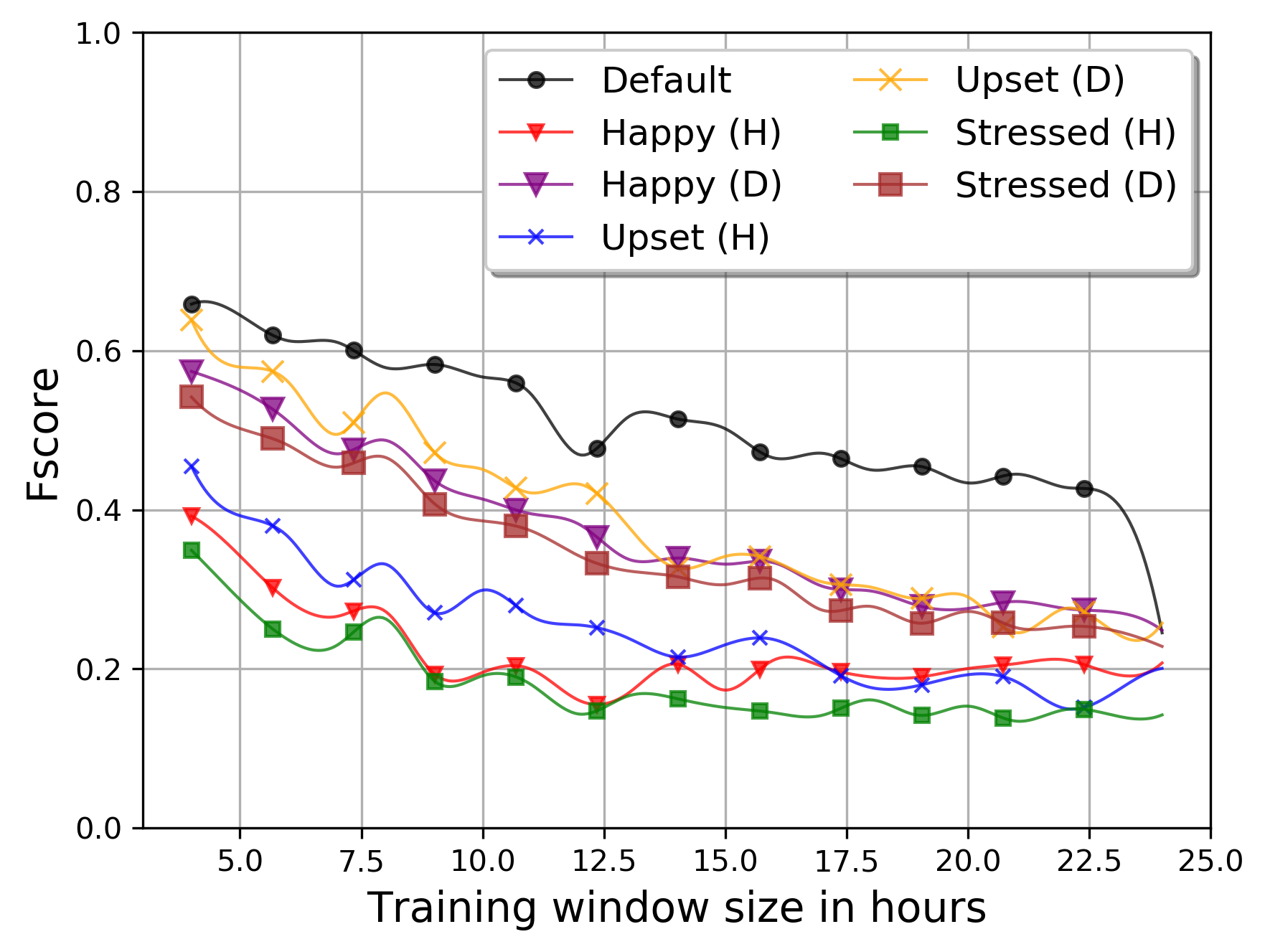}
    \caption{$F$-scores for each experiment using only mood-related samples across different training set sizes.}
    \label{fig:fscores_withmoodonly}
\end{figure}

Thus far, we have evaluated identification performance by removing samples associated with mood. Figure \ref{fig:fscores_withmoodonly} provides $F$-scores when using only the samples corresponding with mood. Here, we see more significant effects that may be associated with happiness, upsetnees, and stress. First, none of these experiments yield results that outperform the use of all samples. We do note, however, that samples associated with mood inferred daily ($D$) perform worse than those inferred by the hour ($H$). Importantly, however, we find that samples related to upsetness yield a $F$-score for four hours of training data comparable to that with the use of all samples. While two of the 19 subjects were never classified as upset, this finding may have several implications. For instance, biometrics for mental health is a developing research area (e.g., \cite{info:doi/10.2196/mhealth.5960}), and considering our results that show that the feeling of being upset may produce identifiable behavior, mHealth treatment services that focus on this particular affect could leverage these findings for personalized patient support.

\subsection{What is the impact of mood on behavior?}
Thus far, our results do not support the notion that mood has a negative impact on identification performance; on the contrary, we find that the inclusion of these samples help to improve identification performance. In this section, we explore changes in phone usage potentially brought on by mood. If variations exist, then we can further claim that mood does cause intra-person variation, but these variations may not degrade performance. 

Table \ref{tbl:chosen_feats} provides the top ten features chosen during feature selection with their normalized frequencies (i.e., how often they were chosen across the different experiments when changing the training set size). These features were selected when running identification experiments using \textit{only} those samples labeled with the respective mood. App\_2 and app\_3 represent the home screen launcher and Gmail apps, respectively. Apps\_ represents no app usage. For features gps\_$i$, $i$ represents the DBSCAN cluster ID. Given these findings, we conclude that there \textit{are} differences in behavior that may be brought on by mood. For instance, not using any apps is selected less frequently when considering mood alone compared to the default configuration, while locking habits become the most frequently selected feature when considering mood. Similarly, differences in feature saliency are observed for calling habits and two locations. An additional finding is that the saliency of audio (e.g., silence, voice, or noise) increases when considering mood. Our results also show that some features remain equally important (i.e., selected at the same frequency) across experiments (e.g., homescreen and Gmail apps). Moreover, we find that the frequency in which activity is selected is only changed when considering upsetness, suggesting that people may physically behave differently when upset. 

\begin{table}[ht]
\centering
\caption{Top ten most selected features with their associated normalized frequencies. These findings suggest variations in behavior that may be dependent on mood. Feature app\_2 corresponds with the home screen launcher, and app\_3 corresponds with the Gmail app. GPS features correspond with DBSCAN clusters.}
\label{tbl:chosen_feats}

\resizebox{\columnwidth}{!}{  
\renewcommand{\arraystretch}{1.3}

\begin{tabular}{l|l|l|l} \hline
\textbf{Default} & \textbf{Happy} & \textbf{Upset} & \textbf{Stressed} \\ \hline
\rowcolor[HTML]{C0C0C0} 
\cellcolor[HTML]{34FF34}(apps\_)	\#: 0.01196 & (lock)	\#: 0.00076 & (lock)	\#: 0.00049 & (lock)	\#: 0.00107 \\\hline
\rowcolor[HTML]{9B9B9B} 
\cellcolor[HTML]{C0C0C0}(lock)	\#: 0.01185 & (audio)	\#: 0.00076 & (audio)	\#: 0.00048 & (audio)	\#: 0.00105 \\\hline
\rowcolor[HTML]{656565} 
\cellcolor[HTML]{DAE8FC}(gps\_0)	\#: 0.01185 & (gps\_1)	\#: 0.00055 & (gps\_1)	\#: 0.00032 & (gps\_1)	\#: 0.00077 \\\hline
\cellcolor[HTML]{9B9B9B}(audio)	\#: 0.01185 & \cellcolor[HTML]{DAE8FC}(gps\_0)	\#: 0.00053 & \cellcolor[HTML]{FFCCC9}{\color[HTML]{000000} (activity)	\#: 0.00031} & \cellcolor[HTML]{FFFC9E}(call)	\#: 0.00074 \\\hline
\rowcolor[HTML]{DAE8FC} 
\cellcolor[HTML]{656565}(gps\_1)	\#: 0.01185 & \cellcolor[HTML]{FFFC9E}(call)	\#: 0.00048 & (gps\_0)	\#: 0.00031 & (gps\_0)	\#: 0.00074 \\\hline
\rowcolor[HTML]{FFCCC9} 
{\color[HTML]{000000} (activity)	\#: 0.01175} & (activity)	\#: 0.00046 & \cellcolor[HTML]{FFFC9E}(call)	\#: 0.00031 & (activity)	\#: 0.00072 \\\hline
\rowcolor[HTML]{34FF34} 
\cellcolor[HTML]{FFFC9E}(call)	\#: 0.01131 & (apps\_)	\#: 0.00039 & (apps\_)	\#: 0.00025 & (apps\_)	\#: 0.00054 \\\hline
\rowcolor[HTML]{FFFFFF} 
{\color[HTML]{000000} (apps\_2)	\#: 0.01035} & {\color[HTML]{000000} (apps\_2)	\#: 0.00036} & {\color[HTML]{000000} (apps\_2)	\#: 0.00025} & {\color[HTML]{000000} (apps\_2)	\#: 0.00048} \\\hline
\rowcolor[HTML]{FFFFFF} 
{\color[HTML]{000000} (apps\_2 3)	\#: 0.00642} & {\color[HTML]{000000} (apps\_2 3)	\#: 0.00021} & {\color[HTML]{000000} (apps\_2 3)	\#: 0.00017} & {\color[HTML]{000000} (apps\_2 3)  \#: 0.00024} \\ \hline

\end{tabular}}
\end{table}

\section{Conclusion} \label{conclusion}
Studies have shown that behavioral patterns can be used to identify an individual. Utilizing this fact, we sought to understand the effect of mood on performance using mobile biometric data. We extracted call, app, GPS, audio, activity, and lock activity features from approximately 60 days of data from 19 subjects in the StudentLife Dataset \cite{wang2014studentlife}. Leveraging the dataset's ecological momentary assessment responses, we explored if being happy, upset, or stressed caused a significant change in identification performance using a random forest classifier, and train/test sizes of four hours to one day. 
When not considering mood, we achieved an $F$-score of approximately 67\% with four hours of training and test data. While this is relatively low, the goal of this study was not to optimize performance, but to measure the impact of mood on performance. Our findings show that by excluding samples associated with a subject's mood, performance worsened, with a best-case $F$-score of 48\%. Thus, while previous works have claimed that behavioral biometrics may be negatively impact by intra-person variations, including mood, our findings show that the variations found in mood do not hinder performance. In fact, our results show that happy, upset, and stressed-related samples account for only 6\%, 6\%, and 10\%, respectively, of the total number of samples. However, when excluding these samples, performance significantly worsens by 28.4\%. 

\begin{figure}[ht]
    \centering
     \includegraphics[width=\columnwidth]{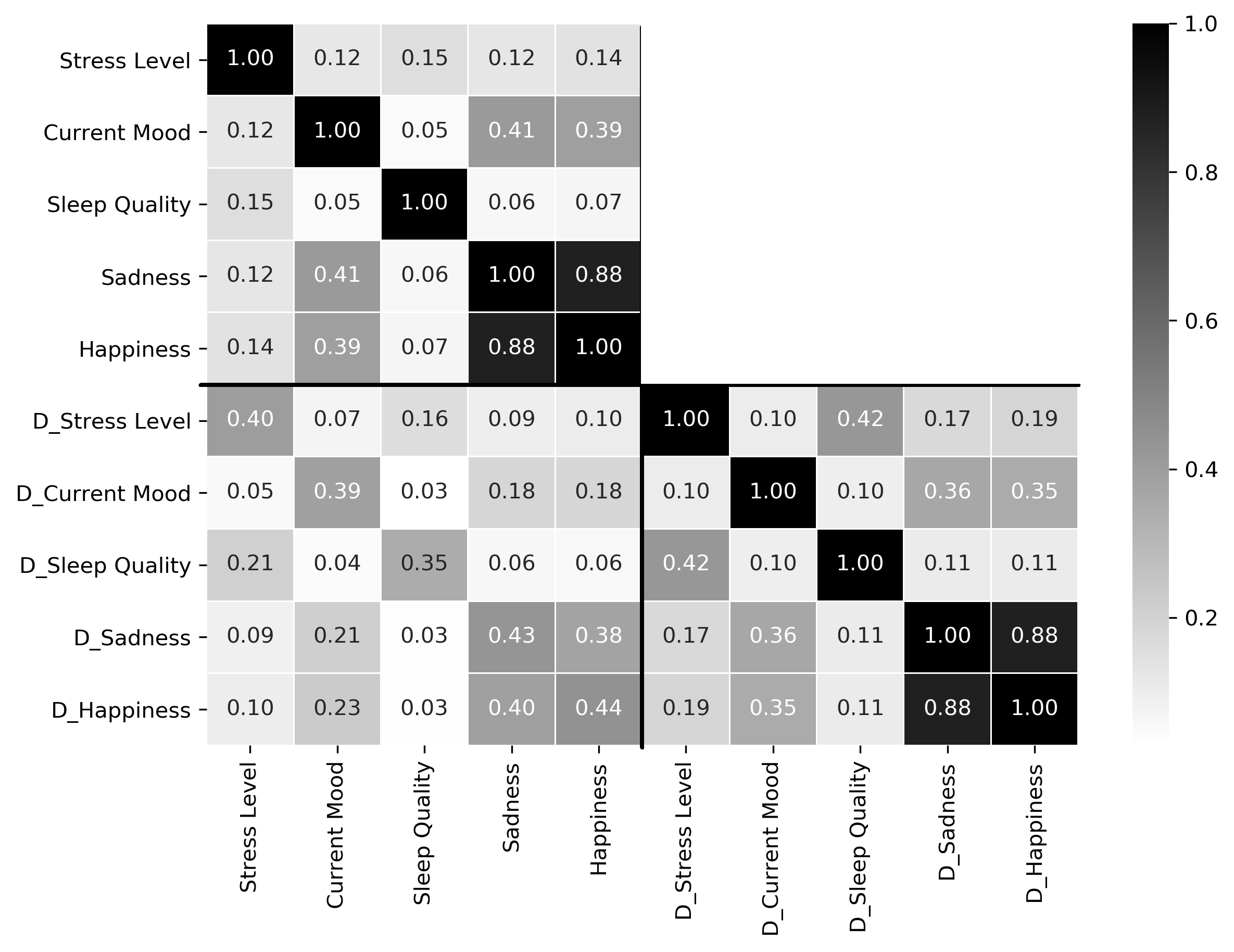}
    \caption{Pearson's correlation coefficients on EMA responses inferred hourly and daily (latter indicated by D\_). }
    \label{fig:ema_correlations}
\end{figure}

Finally, we note the limitations of this work and directions for future work. First, we highlight that the dataset used is small, making it difficult to draw concrete conclusions. This leads to an important area of future work. When preparing our experiments, we did not find any publicly available datasets that included mobile device usage data with annotated mood. The StudentLife dataset did, however, contain this data, and we chose 19 of the subjects from this dataset based on empirical analysis to determine which subjects had the most
representation in the dataset in regard to data availability and EMA responses. This was necessary
to avoid highly imbalanced classes. However, to obtain statistically significant results, a larger dataset is necessary.

A second limitation is the skewed distribution of the data for the different moods, which may have impacted the reported drop in accuracy due to lesser data points to learn from. Considering this, our analysis evaluated this potential problem by removing mood-related samples and considering only mood-related samples. We further explored if our results were based solely on the lack of data points by using feature selection, finding that feature importance changed for each mood. Thus, while the number
of data points affects performance, the fact that the salient features change is an indication that
different moods impact mobile biometric performance in different ways. Nonetheless, this observation is also related to the availability of data which allows sufficient exploration of this problem.

An additional limitation is the assumptions made throughout that essentially drive the experimental set-up and drawn conclusions. Many of these assumptions rely on the accuracy of perceived mood gathered by ecological momentary assessments, or self-reports. However, certain correlations of mood were not all supported by previous literature. For instance, poor sleep is known to correlate with decreased allostatic load \cite{hamilton2007sleep, Chen2006}; however, in Figure \ref{fig:ema_correlations}, we see a positive correlation between these two factors. Recall that in this paper, the numerical representation for sleep quality increases as sleep quality worsens, while the numerical label for stress increases as the user feels less stressed. These correlations may be due to the subjectivity of self-report, and the subjects' inability to accurately express how they feel. There were also few EMA responses, impacting our experimental design to infer mood when this data was unavailable. This may have also played a role in the observed correlations found in Figure \ref{fig:ema_correlations}. Nonetheless, while our results may be difficult to generalize to the general public, we have attempted to provide insight into the variability of mood on mobile biometrics. In future work, we will explore additional potential ``mobile occlusions'' that may be problematic for mobile biometric systems, including additional moods and emotions. 

{\small
\bibliographystyle{plain}
\bibliography{references}
}

\end{document}